\documentclass[aps,prl,twocolumn,superscriptaddress,showpacs,floatfix,longbibliography]{revtex4-2}
\usepackage{amsmath,amssymb,amsfonts,float,graphics,epsfig,epstopdf,color,verbatim,tabularx,bm,multirow,appendix,wasysym}
\usepackage[utf8]{inputenc}
\usepackage[T1]{fontenc}
\usepackage{xcolor}
\usepackage{dsfont}
\usepackage{textcomp}
\usepackage{yfonts}
\usepackage{bm}
\usepackage{subfigure}
\usepackage{mathrsfs}
\usepackage{graphicx}
\usepackage{verbatim}
\usepackage{hyperref}
\usepackage{multirow}
\usepackage{braket}
\usepackage[normalem]{ulem}
\usepackage{tikz}
	\usetikzlibrary{calc}
	\usetikzlibrary{shapes.multipart}

\usepackage{bbm}

\newcommand{\comments}[1]{}

\def\H{\mathcal{H}}

\newcommand{\tcb}{\textcolor{black}}

\newcommand{\stkout}[1]{\ifmmode\text{\sout{\ensuremath{#1}}}\else\sout{#1}\fi}

\newcommand*{\red}{\textcolor{black}}

\usepackage{natbib} 
\makeatletter
\def\l@subsubsection#1#2{}
\makeatother

\begin{document}
\title{Sampling reduced density matrix to extract fine levels of entanglement spectrum and restore entanglement Hamiltonian}

\author{Bin-Bin Mao}
\affiliation{Department of Physics, School of Science and Research Center for Industries of the Future, Westlake University, Hangzhou 310030,  China}
\affiliation{School of Foundational Education, University of Health and Rehabilitation Sciences, Qingdao 266000, China}

\author{Yi-Ming Ding}
\affiliation{State Key Laboratory of Surface Physics and Department of Physics, Fudan University, Shanghai 200438, China}
\affiliation{Department of Physics, School of Science and Research Center for Industries of the Future, Westlake University, Hangzhou 310030,  China}

\author{Zhe Wang}
\affiliation{Department of Physics, School of Science and Research Center for Industries of the Future, Westlake University, Hangzhou 310030,  China}
\affiliation{Institute of Natural Sciences, Westlake Institute for Advanced Study, Hangzhou 310024, China}

\author{Shijie Hu}
\email{shijiehu@csrc.ac.cn}
\affiliation{Beijing Computational Science Research Center, Beijing 100193, China}
\affiliation{Department of Physics, Beijing Normal University, Beijing 100875, China}

\author{Zheng Yan}
\email{zhengyan@westlake.edu.cn}
\affiliation{Department of Physics, School of Science and Research Center for Industries of the Future, Westlake University, Hangzhou 310030,  China}
\affiliation{Institute of Natural Sciences, Westlake Institute for Advanced Study, Hangzhou 310024, China}

\begin{abstract}
The reduced density matrix (RDM) plays a key role in quantum entanglement and measurement, as it allows the extraction of almost all physical quantities related to the reduced degrees of freedom. However, restricted by the degrees of freedom in the environment, the total system size is often limited, let alone the subsystem. To address this challenge, we propose a quantum Monte Carlo scheme with a low technical barrier, enabling precise extraction of the RDM. To demonstrate the power of the method, we present the fine levels of the entanglement spectrum (ES), which is the logarithmic eigenvalues of the RDM. We clearly show the ES for a $1$D ladder with a long entangled boundary, and that for the $2$D Heisenberg model with a tower of states. Furthermore, we put forward an efficient way to restore the entanglement Hamiltonian in operator-form from the sampled RDM data. Our simulation results, utilizing unprecedentedly large system sizes, establish a practical computational framework for determining entanglement quantities based on the RDM, such as the ES, particularly in scenarios where the environment has a huge number of degrees of freedom.
\end{abstract}

\maketitle

\vspace{\baselineskip}
\noindent{\bf Introduction}\\
Quantum information and condensed matter physics have been increasingly cross-fertilizing each other in recent years~\cite{Amico2008entanglement,laflorencie2016quantum}. Within this trend, quantum entanglement was proposed to detect the field-theoretical and topological properties of quantum many-body systems~\cite{vidal2003entanglement,Korepin2004universality,Kitaev2006,Levin2006}. For example, it usually offers the direct connection to the conformal field theory (CFT) and categorical description~\cite{Calabrese2008entangle,Fradkin2006entangle,Nussinov2006,Nussinov2009,CASINI2007,JiPRR2019,ji2019categorical,kong2020algebraic,XCWu2020,JRZhao2020,XCWu2021}. 
Beyond the entanglement entropy (EE), Li and Haldane proposed that the entanglement spectrum (ES) is a more fundamental entanglement characteristic to quantify the intrinsic information of many-body systems~\cite{Li2008entangle,Thomale2010nonlocal,Poilblanc2010entanglement,schliemann2012entanglement,laflorencie2016quantum}. They also suggested a profound correspondence between the ES of an entangled region with the energy spectra on the related virtual edges, which is the so-called Li-Haldane conjecture. 
Thereafter, low-lying ES has been widely used as a fingerprint to investigate CFT and topology in highly entangled quantum systems~\cite{Pollmann2010entangle,Fidkowski2010,Yao2010,XLQi2012,Canovi2014,LuitzPRL2014,LuitzPRB2014,LuitzIOP2014,Chung2014entanglement,Pichler2016,cirac2011entanglement,Stoj2020,guo2021entanglement,Grover2013,Assaad2014,Assaad2015,Parisen2018,yu2021conformal,RoosePrb2021,RoosePrb2022,yan2021extract,liu2023probing,wu2023classical,song2023different}.

However, most of the ES studies so far have focused on (quasi) 1D systems numerically, due to the exponential growth of computation complexity and memory cost. The existing numerical methods such as the exact diagonalization (ED) and the density matrix renormalization group (DMRG) have significant limitations for approaching large entangling region. 
We find it noteworthy that the quantum Monte Carlo (QMC) serves as a powerful tool for studying large size and higher dimensional open quantum many-body systems~\cite{zcai2014,zyan2018}. 
QMC has been refined to extract the entanglement spectral function combined with numerical analytic continuation~\cite{zyan2021entanglement,wu2023classical,song2023different,liu2023probing,Assaad2014,Assaad2015,Parisen2018,SHAO2023,sandvik1998stochastic,Beach2004,sandvik2016constrained,OFS2008using,Shao2017nearly,ZY2020,Zhou2021amplitude}. Though the method overcomes the exponential wall problem and can obtain large-scale entanglement spectral functions, it suffers from limited
precision and fails to distinguish the fine levels while the level structure always carries important information of CFT and topology. Moreover, the spectral function is also different from the spectrum, which we will further compare to demonstrate that their characteristics sometimes may be totally different.
The other way to extract ES through QMC is to reconstruct it according to the knowledge of many integer R\'enyi entropies because integer R\'enyi entropies can be calculated by QMC~\cite{Chung2014entanglement}. The difficulty of this scheme is that the calculation of $n$-th R\'enyi entropy itself is not an easy task especially for higher $n$, while the higher $n$-th entropy contains more information of low-lying levels.

All in all, the reduced density matrix (RDM) as a generator of almost all the entanglement quantities, takes a key role in quantum many-body physics but is limited seriously by the huge degree of freedom. From the RDM, almost all physical quantities in the reduced degree of freedom can be extracted.
Once the reduced density matrix can be simulated in large size, the game can be changed.
Therefore, a practical scheme to extract fine RDM with reduced computation complexity is urgent to be proposed. In this paper, we develop a protocol which can efficiently compute the RDM via quantum Monte Carlo simulation for the entangling region with long boundaries and in higher dimensions. The large-scale ES was selected to verify the precision of the method for two reasons: On the one hand, unlike EE, ES needs to deal with a large amount of degrees of freedom; on the other hand, the requirement of accuracy is extremely high because ES is the logarithm of RDM.
To demonstrate the strength of our method, we carefully investigated a Heisenberg ladder, selecting one chain as the entangling region.
Compared with previous work studying the same system~\cite{Poilblanc2010entanglement,lauchli2012entanglement,cirac2011entanglement}, {we can not only address a much larger size, but also} clarify some misunderstandings in past work~\cite{zyan2021entanglement,Poilblanc2010entanglement} through our large-scale calculations. 
Furthermore, we study a 2D Heisenberg model to reveal the nontrivial continuous symmetry breaking phase using the structure of tower of states in the ES.
{Except obtaining the ES numerically, we put forward an efficient way to restore the entanglement Hamiltonian in operator-form from the \tcb{RDM} data.}
It is important to emphasize that our method is not limited to these particular systems, as it can be widely applied to any models which can be simulated via QMC.

\vspace{\baselineskip}
\noindent{\bf Results}\\
\noindent\textit{\color{blue}Method.-}
In a quantum many-body system, the ES of a subsystem (entangling region) $A$ with the rest of the system $B$ (environment) is constructed via the reduced density matrix. The RDM is defined as the partial trace of the total density matrix $\rho$ over a complete basis of $B$, i.e. $\rho_{A}=\mathrm{Tr}_{B}(\rho)$. The RDM $\rho_{A}$ can be interpreted as an effective {thermal} density matrix $e^{-\mathcal{H}_E}$ \tcb{at temperature $T=1$} through an entanglement Hamiltonian $\mathcal{H}_E$,  {namely $\H_E=-\mathrm{ln}(\rho_A)$}. The spectrum of the entanglement Hamiltonian is usually denoted as the ES.

Due to the exponential growth of the computation complexity and the memory cost, it is intractable for the existing numerical methods such as the ED and the DMRG to approach entangling regions with long boundaries and higher dimensions. Because of the finite computer memory, these numerical methods are limited to short boundary only. Another way is to extract the low-lying ES through the quantum Monte Carlo simulation combined with stochastic analytic continuation (SAC), both in bosonic~\cite{zyan2021entanglement,liu2023probing,li2023relevant} and fermionic systems~\cite{Assaad2014,Assaad2015,Parisen2018}. Especially in bosonic systems, the ES of spin ladders can be obtained even under a long length~\cite{zyan2021entanglement}. However, the ES obtained in this way can not show the fine levels but only the dispersion and weight of the spectral function. Nevertheless, the refined structure of ES, e.g., degeneracy, is important for identifying the CFT and topology.

The solution comes from QMC $+$ ED: Tracing the environment via QMC and obtaining the exact low-lying levels through ED. The RDM element $|C_A\rangle\langle C'_A|$ can be written in the path integral language as
\begin{equation}
{\rho_A}_{C_A, C'_A}\propto\sum_{\{C_B\}}\langle C_A, C_B|e^{-\beta H}|C'_A, C_B\rangle, ~{\beta\rightarrow\infty},
\label{eq1}
\end{equation}
where $C_A,~C'_A$ are the configurations of the RDM, and $C_B$ is the configurations of environment $B$. 

\red{Commonly, we aim to simulate a partition function $\sum_{\{C\}}\langle C|e^{-\beta H}|C\rangle$, where $\{C\}$ \tcb{represents a complete, orthogonal basis set} in Hilbert space. The configurations $C$ in the bra $\langle ... |$ and ket $|...\rangle$ are same, which means \tcb{that} the boundary condition of \tcb{the} imaginary time \tcb{in} the path integral is periodic. If we want to calculate \tcb{matrix elements} $\langle C|e^{-\beta H}|C'\rangle$ \tcb{in} the basis \tcb{set} $\{C\}$, we can achieve this by setting an open boundary condition \tcb{for the} imaginary time \tcb{in} the path integral, \tcb{allowing $C$ and $C'$ to differ}.}

\begin{figure}
    \centering
    \includegraphics[width=0.98\columnwidth]{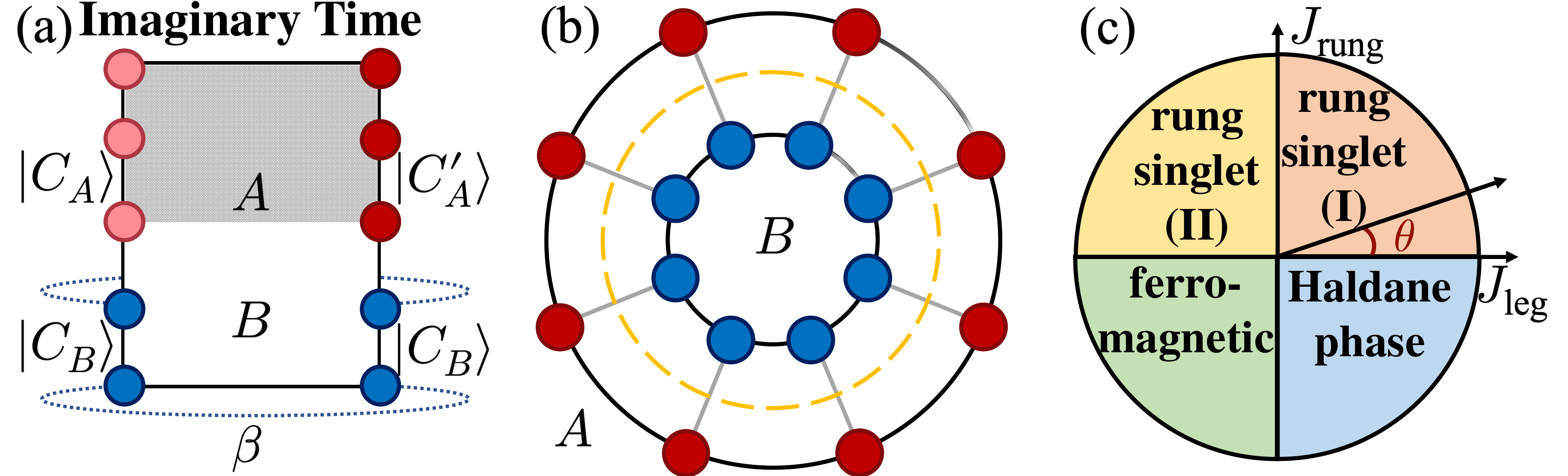}
    \caption{The schematic diagram of sampling reduced density matrix and the characters of Heisenberg ladder. (a) A path integral configuration of a reduced density matrix. The length of imaginary time is $\beta$, and the time boundary of $B$/$A$ is periodic/open; (b) Lattice of the Spin-1/2 Heisenberg ladder from an overhead view, with spins in $A$ and $B$ depicted by red and blue circles respectively; (c) Its phase diagram by denoting $J_{\text{leg}}=\cos\theta$ and $J_{\text{rung}}=\sin\theta$.}
    \label{Fig1}
\end{figure}

Following the above understanding, the Eq. (\ref{eq1}) can be treated as a general partition function under a special boundary condition along the imaginary time axis as Fig.~\ref{Fig1}(a) shows, which is much more convenient to be simulated by quantum Monte Carlo. We deal with this special partition function in Eq. (\ref{eq1}) in the framework of stochastic series expansion (SSE) method~\cite{Sandvik1991,Sandvik1999,Syljuaasen2002,ZY2019,yan2020improved}. Of course it can be simulated by other path integral QMC~\cite{suzuki1977monte,PhysRevB.26.5033,suzuki1976relationship,PhysRevE.66.066110,huang2020worm,fan2023clock}. The only difference compared with the normal SSE is that we open the boundary of the imaginary time in the region $A$ and keep the periodic boundary condition to that of the environment $B$.

The value of each element in the RDM can be approached with the frequency of the samplings. It means that the ${\rho_A}_{C_A, C'_A}$ is proportional to $N_{C_A, C'_A}/N_{\text{total}}$, where the $N_{C_A, C'_A}$ is the amount of samplings that the upper/below imaginary time boundary configuration is $C_A/C'_A$ and $N_{\text{total}}$ is the total amount of samplings. It's worth noting that the sign should be considered if the weight is not positive.

\noindent\textit{\color{blue}Example 1: Antiferromagnetic (AFM) Heisenberg ladder.-} 
To demonstrate the power of our method, we compute the ES of two-leg Heisenberg ladder where the entangled boundary splits the ladder into two chains as shown in Fig.~\ref{Fig1}(b). The ES of this model has been carefully studied via ED for small sizes in previous work~\cite{Poilblanc2010entanglement} and it has a well-known phase diagram (See Fig.~\ref{Fig1}(c)). 
The spins on the ladder are coupled through the nearest neighbor interactions, where $J_{\text{leg}}$ denotes the intra-chain strength along the leg and $J_{\text{rung}}$ denotes the inter-chain strength on the rung. We first simulate with $J_{\text{leg}}=1$ and $J_{\text{rung}}=1.732$ ($\theta=\pi/3$) at $\beta=100$. The reason for this choice of parameters is that $\beta=100$ is a temperature low enough for a gapped rung singlet (I) phase which has been studied with ED in Ref.~\cite{Poilblanc2010entanglement}, therefore it is convenient for us to compare the results.

\begin{figure}[htp]
\centering
\includegraphics[width=0.98\columnwidth]{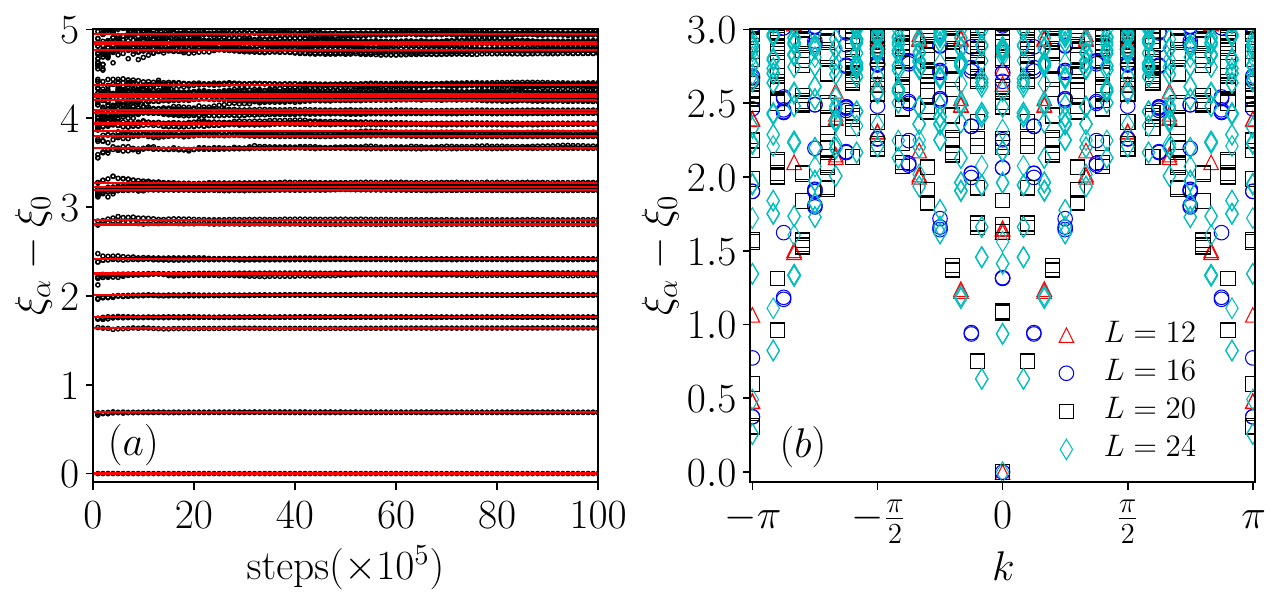}
\caption{Comparison of entanglement spectra using different methods. (a) The comparison of low lying entanglement excitation spectra obtained by pure ED and QMC+ED with different Monte Carlo steps. The red lines are the results obtained by pure ED and the black dots are the results obtained by QMC+ED. Here we choose $L=8$. (b) Entanglement excitation spectra versus total momenta $k$ in the chain direction for $\theta=\pi/3$.}
\label{Fig_MC_ED}
\end{figure}

\begin{figure}[htb] 
 \begin{center}
    \includegraphics[width=0.98\columnwidth]{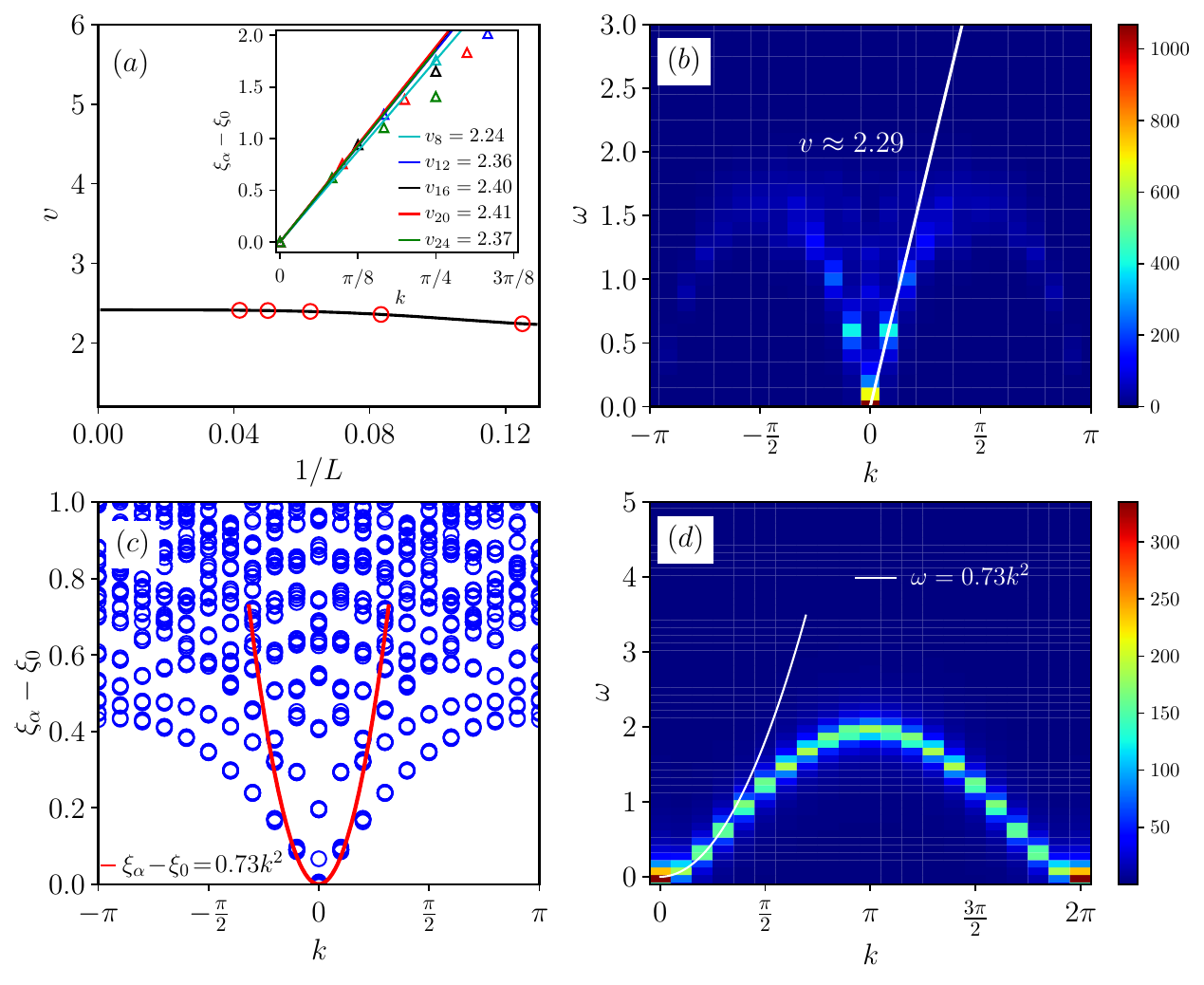}  
\end{center}
    \caption{ {
      Extrapolation and spectral analysis of entanglement spectrum in the spin ladder system. (a) Extrapolation of velocity $v$ as a function of $1/L$. The inset shows the fitting of velocity $v_{L}$ based on the numerical results for various size $L$.  (b) Spectral function of the system with size $L=24$ and its corresponding velocity $v$ fitting.
        (c) Entanglement spectrum versus total momenta $k$ in the chain direction for $\theta=2\pi/3$ with the system size $L=20$. The red line is the  quadratic fitting curve in the entanglement spectrum, given by $\omega=0.73k^2$.
(d) The entanglement spectral function of FM Heisenberg ladder with $L=20$. The white line is the quadratic fitting curve near the gapless point.  }
        }
        \label{Fig_AFM_FM}
\end{figure}
Firstly, we present a comparison of the spectrum obtained through QMC+ED and directly through ED with system size $L=8$, as illustrated in Fig.~\ref{Fig_MC_ED}(a). It is evident that the spectrum obtained via QMC+ED approaches that of ED as the number of samplings increases, and higher levels require a greater number of sampling iterations. It is easy to be understood because the QMC of course samples the lower energy configurations with heavier weight preferentially, thus the low-lying levels converge first.

After demonstrating the feasibility of this method, we want to answer some physical questions. For example, there is a discrepancy for the ES in the rung singlet (I) phase of the Heisenberg ladder in previous works~\cite{Poilblanc2010entanglement,zyan2021entanglement}. The ES here is expected to bear the low-energy CFT structure, i.e., the ground level of ES, $\xi_0$, will scale as $\xi_0/L=e_0+d_1/L^2+\tcb{O}(1/L^3)$ where the $d_1=\pi c v/6$ according to the CFT predication with the central charge $c=1$ and $v$ is the velocity of ES near the gapless point~\cite{Poilblanc2010entanglement}, \tcb{i.e.,} the Cloiseaux-Pearson spectrum of the quantum spin chain, \tcb{$v|\sin k|$}~\cite{Cloizeaux1962}. The fitting for the fine levels of the ES by ED at $L=14$ (its largest size) gave $v\sim 2.36$ while the fitting for the rough spectral function of entanglement by QMC combined with SAC at $L=100$ pointed to a double value $v\sim 4.58$~\cite{zyan2021entanglement}. \red{Therefore, the later paper \tcb{attributed the different results} to the finite size effect.}
Since our method avoids the limitation of the environment size, we can calculate larger sizes to see how the $v$ scales with $L$. If the ES has obvious finite size effect, the $v$ certainly will increase following the size. 


The spectra of the system with various, but larger system sizes obtained by QMC+ED is shown in Fig.~\ref{Fig_MC_ED}(b), and the results are consistent with the ED results show in Ref.~\cite{Poilblanc2010entanglement}. 
It demonstrates the efficiency of our method for larger systems, so that we further inspect the finite size effect here.
In Fig.~\ref{Fig_AFM_FM}(a), we draw a fitting line for the velocity $v$ of different size $L$. In the inset of Fig.~\ref{Fig_AFM_FM}(a), 
we fit the velocities of different sizes by the first two data points. All the lines are close to each other, representing that their velocities are similar.
The fitting line in Fig.~\ref{Fig_AFM_FM}(a) turns out that the velocity is almost unchanged with size $L$ and convergent to $\sim 2.41$, which is inconsistent to the result of spectral function by QMC+SAC where $v\sim 4.58$\tcb{~\cite{zyan2021entanglement}}.

In order to show the spectral property of the system and avoid \tcb{any puzzling inconsistency} between the ES and entanglement spectral function, we also \tcb{examine} the spectral function of entanglement. For a physical observable denoted by $\mathcal{O}$, the spectral function $S(\omega)$ \tcb{in general} can be written as
\begin{equation}
S(\omega) = \frac{1}{\pi}\sum_{m,n} e^{-\beta E_n}\left|\left\langle n \right| \mathcal{O} \left| m \right\rangle\right|^2 \delta(\omega - \left[{E_m-E_n}\right])
\label{sf}
\end{equation}
where $|n (m)\rangle$ and $E_{n(m)}$ are the \tcb{eigenstates and eigenvalues of the given Hamiltonian, respectively.} \red{This definition \tcb{applies to both the} original Hamiltonian \tcb{and the} entanglement Hamiltonian. \tcb{In this work, we} focus on the spectral function of \tcb{the} entanglement Hamiltonian.}
In the following, we choose $\mathcal{O}$ as the spin operator $S_k^z$, and fix $|n\rangle$ as the ground state, i.e. \tcb{$E_n \equiv E_0$ is the ground-state} energy (it is exact when $\beta\rightarrow\infty$). The choice is same as the Ref.~\cite{zyan2021entanglement} which is convenient for comparisons. \tcb{It is} worth noting that the Hamiltonian and energy level in \tcb{Eq.~(\ref{sf})} should be replaced as entanglement Hamiltonian and ES level when we talk about the ES.

Fig.~\ref{Fig_AFM_FM}(b) shows the spectral function of the system with size $L=24$ and we also give the velocity $v\sim 2.29$. It reveals that the discrepancy between the ES and entanglement spectral function is not because of the difference of the definition. We find the reason may be the settings of constant missing a factor ``$2$" in the Ref.~\cite{zyan2021entanglement}.

\noindent\textit{\color{blue} Example 2: Ferromagnetic (FM) Heisenberg ladder.-} Furthermore, we simulate the case with ferromagnetic $J_{\text{leg}}=-1$ and antiferromagnetic $J_{\text{rung}}=1.732$ at $\beta=100$ on the two-leg Heisenberg ladder for $L=8, 12, 16, 20$ and compare the results with those in Ref.~\cite{Poilblanc2010entanglement}, in which case the ladder is in the gapped rung singlet (II) phase.  

According to Refs.~\cite{Fogedby1980,Haldane1982,zyan2021entanglement}, the lowest-energy excitations at \tcb{momentum} $k=2\pi/L$ corresponds to the one-magnon state and behaves as \tcb{$E=2 J_{\text{eff}}\ {\sin}^2(k/2)$}, where $J_{\text{eff}}$ is an effective chain coupling. The results in Ref.~\cite{Poilblanc2010entanglement} did not show the quadratic behavior due to the small system size. Other studies explained this non-quadratic dispersion via long-range interactions~\cite{lauchli2012entanglement,cirac2011entanglement,laflorencie2016quantum}.
Meanwhile, the QMC+SAC simlation at $L=100$ in the Ref.~\cite{zyan2021entanglement} shows the spectral function of entanglement is indeed a quadratic dispersion, which further \tcb{implied} the finite size effect of ES here.

In order to check whether the loss of quadratic dispersion is because of the finite size effect, we further try to calculate the ES and its spectral function for larger sizes. Fig.~\ref{Fig_AFM_FM}(c) shows the results with $L=20$, in which there is no obvious quadratic dispersion near $k=0$. Actually, we \tcb{have not} seen an obvious change of the dispersion from the linear to the quadratic when increasing the size. 

In fact, the answer of the discrepancy comes from the difference between the ES and entanglement spectral function instead of the finite size effect. The results of the FM Heisenberg ladder indicate that the \tcb{energy} spectrum and \tcb{ordinary} spectral function exhibit distinct behaviors at low energy levels. As it is shown in Fig.~\ref{Fig_AFM_FM} (d), the ES in $L=20$ does not show a quadratic dispersion while the spectral function does. \red{It is worth noting that we measure the excitation of $S_k^z$ in the spectral function [Eq.(\ref{sf})], which is actually a one-magnon dispersion. It means the excitation of one-magnon is indeed quadratic here even in small size.} We try to draw the curve of the spectral function in the ES levels as the red line shown in Fig.~\ref{Fig_AFM_FM} (c), which further demonstrates that the spectral function goes up quickly beyond the lowest levels of ES. \red{We find that some previous studies in quantum magnetism show, in a pure 1D FM Heisenberg chain, its one-magnon excitation is also higher than the two-magnon bound state~\cite{Bethe1931,Karabach1997}. It supports our conclusion that the one-magnon is not the lowest excitation in the FM chain.}
Thus, the spectral function behaves differently from the low-lying ES in the ferromagnetic case.


\noindent\textit{\color{blue} Example 3: 2D Heisenberg model.-}
For a $d\geq 2$ dimensional quantum $O(N)$ model, its ground state spontaneously beaks the continuous symmetry and is labeled by N\'eel order. Meanwhile, its energy spectrum has the characteristic of tower of states (TOS) structure. The energies of the TOS in finite size are~\cite{hasenfratz1993finite,Deng2023improved} 
\begin{equation}
E_S(L)-E_0(L)=\frac{S(S+N-2)}{2\chi_{\perp}L^d}
\label{eq3}
\end{equation}
where $S$ is the total spin momentum of the system, $\chi_{\perp}$ is the transverse susceptibility in the thermodynamic limit. Except the energy spectrum~\cite{wietek2017studying}, the ES is also expected having TOS when the ground state is continuous symmetry breaking~\cite{Alba2013entanglement}. 

\begin{figure}[htb]
\centering
\includegraphics[width=\columnwidth]{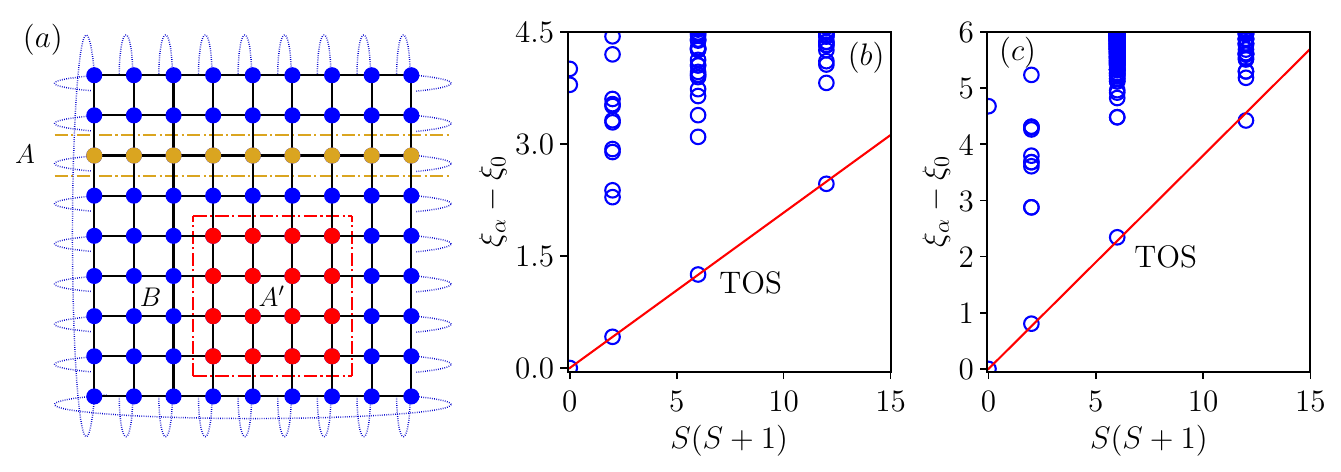}
\caption{The entanglement spectrum of entanglement Hamiltonian in 2D system. (a) The AFM Heisenberg model on the square lattice. 
The dashed lines are used to illustrate the bipartition into two subsystems. The yellow dashed lines illustrate the cutting method that $A$ is a ring and the red dashed lines illustrate the method that $A^\prime$ is a block. For each cutting method, the other part is denoted as $B$. We consider the square lattice model with size $20\times 20$. We show the entanglement spectrum of EH corresponding to (b) $A$ with length $L=20$ and (c) $A^\prime$ with size $4\times 4$. 
The TOS levels are connected by a red line. All the data are calculated in the total $S^z=0$ sector.}
\label{fig:TOS}
\end{figure}

We then calculate a periodic boundary condition (PBC) $20\times 20$ square lattice AFM Heisenberg model with two different cut-geometries of $A$. If we choose the $A$ as a chain as the yellow dots ($A$ region) in Fig.~\ref{fig:TOS}(a) displays, it presents a linear dispersion ES reflecting the N\'eel order of this AFM model, as shown in Fig.\ref{fig:TOS}(b). Because the Heisenberg model belongs to quantum $O(3)$ models, thus the ES levels of TOS is proportional to $S(S+1)$, where the lowest edge of the ES contains the TOS levels~\cite{Kolley2013entanglement,Alba2013entanglement,laflorencie2016quantum}.

Meanwhile, if we set $A$ as a $4\times 4$ region as the red dots ($A'$ region) in Fig.~\ref{fig:TOS}(a), the ES also shows a TOS character as shown in Fig.\ref{fig:TOS}(c). Usually, the TOS holds in cornerless cutting cases, this result reveals that the cornered cutting affects TOS little, at least, in the Heisenberg model. For this $L=4$ square region, we obtain that the velocity of the TOS fitting line is about $0.38$. According to the Eq.(\ref{eq3}), we gain the $\chi_{\perp}\sim0.08$. The $\chi_{\perp}$ here is basically consistent with the QMC result ($\sim0.07$~\cite{Deng2023improved}) of square lattice AFM Heisenberg model in thermaldynamic limit obtained through size extrapolation. (Although the ES should not be exactly same as the energy spectrum, they are always similar according to the Li-Haldane-Poilblanc conjecture.)


\noindent\textit{\color{blue} Restore entanglement Hamiltonian (EH).-}
There is no general method for restoring the EH in operator-form directly from the original Hamiltonian.
In this work, we propose a two-step strategy to determine the EH, spanned by a specific set of orthogonal operators $\{\mathcal{O}_n\}$, without any approximations or a priori assumptions.
For a uniform lattice with $L$ sites, the operator $\mathcal{O}_n$ is defined as a product of orthogonal operators $\mathcal{Q}^p_i$ at each lattice site, i.e., $\mathcal{O}_n = \otimes^L_{i=1} \mathcal{Q}^{k^{n}_i}_i$, where $k^{n}_i$ indexes the operator at site $i$.
For the Hilbert space at each lattice site, with dimension of $\mathcal{N}$, it is natural to set the operator $\mathcal{Q}^p_i$ as one of generators of SU($\mathcal{N}$) group, including the identity operator as an adjoint.
Thus, $p$ ranges from $1$ to $\mathcal{N}^2$, and the operator $\mathcal{O}_n$ has $\mathcal{N}^{2L}$ possible configurations.
As an example, in the AFM Heisenberg ladder, where $\mathcal{N}=2$, we identify the operators as follows: $\mathcal{Q}^0_i = \mathbbm{1}$,  $\mathcal{Q}^1_i = 2 S^z_i$, $\mathcal{Q}^2_i = \sqrt{2} S^+_i$, and $\mathcal{Q}^3_i = \sqrt{2} S^-_i$.

In the two-step strategy, the first step is to extract the numerical RDM simulated by QMC.
The EH can then be calculated from the RDM using the relation $\H_E=-\mathrm{ln}(\rho_A)$, which can be expressed in operator-form as $\H_E = \sum_n J_n \mathcal{O}_n$.
Here the $J_n$ means the coefficient of the operator $\mathcal{O}_n$ in the EH.
The next step involves determining the coefficient $J_n$.
The numerical EH matrix can be projected into the space of orthogonal operators \tcb{via} the equation $J_n = \mathrm{Tr}(\H_E \mathcal{O}_n) / \mathcal{N}^L$, due to $\mathrm{Tr} (\mathcal{Q}^p_i \mathcal{Q}^q_i) = \delta_{p,q} \mathcal{N}$.
In summary, our scheme is both exact and straightforward, allowing the EH to be restored in operator-form.

To demonstrate the power of our method, we restore the EH from the RDM of Heisenberg ladder (our example 1). After consideration of the symmetry, the EH of this system should be a Heisenberg chain with possible long-range interactions~\cite{wu2023classical}. According to the 2nd order perturbation calculation~\cite{Zhuwei2019}, the EH of this ladder in the limit $J_{\mathrm{leg}} \ll J_{\mathrm{rung}}$ is 
\begin{equation}
\tcb{\mathcal{H}_E}\approx J_1\sum_{i}S_i\cdot S_{i+1} + J_2\sum_{i}S_i\cdot S_{i+2}
\end{equation}
where \tcb{the couplings are} $J_1=2\frac{J_{\mathrm{leg}}}{J_{\mathrm{rung}}}$ and $J_2=-\frac{1}{2}\left(\frac{J_{\mathrm{leg}}}{J_{\mathrm{rung}}}\right)^2$. 

\begin{table}[H]
\begin{centering}
\begin{tabular}{|c|c|c|c|c|c|}
\hline 
$J_{\mathrm{rung}}$ &  & $L=8$ & $L=12$ & $L=16$ & Perturbation\tabularnewline
\hline 
\multirow{2}{*}{50} & $J_1$ & 0.04015(3) & 0.03997(1) & 0.0407(2) & 0.04\tabularnewline
\cline{2-6} \cline{3-6} \cline{4-6} \cline{5-6} \cline{6-6} 
 & $J_2$ & -0.00021(2) & -0.00024(2) & -0.0002(1) & -0.0002\tabularnewline
\hline 
\multirow{2}{*}{2} & $J_1$ & 0.9901(2) & 0.9899(4) & 0.9893(5) & 1\tabularnewline
\cline{2-6} \cline{3-6} \cline{4-6} \cline{5-6} \cline{6-6} 
 & $J_2$ & -0.1265(6) & -0.1253(4) & -0.1250(3) & -0.125\tabularnewline
\hline 
\multirow{2}{*}{0.5} & $J_1$ & 3.069(1) & 2.860(1) & 2.549(2) & --\tabularnewline
\cline{2-6} \cline{3-6} \cline{4-6} \cline{5-6} \cline{6-6} 
 & $J_2$ & -1.575(2) & -1.329(3) & -1.244(3) & --\tabularnewline
\hline 
\end{tabular}
\par\end{centering}
\caption{The couplings restored from the sampled RDM for the system with various $J_{\mathrm{rung}}$ for the Heisenberg ladder.}
\label{table1}
\end{table}

We simulate the AFM Heisenberg ladder at $J_{\mathrm{rung}}=50,~2,~0.5$ with fixed $J_{\mathrm{leg}}=1$ to obtain the RDM and then restore the related couplings according to the above scheme. \tcb{The estimates for} $J_1$ and $J_2$ are displayed in the Table~\ref{table1}. 
\tcb{The case with a} large $J_{\mathrm{rung}}=50$ fits well with the perturbation, while $J_{\mathrm{rung}}=2$ \tcb{also provides a good match}. \tcb{However, when} $J_{\mathrm{rung}}=0.5<J_{\mathrm{leg}}=1$, the perturbation theory no longer holds.
In such a case, we find that $|J_2|$ goes much closer to $|J_1|$, \tcb{indicating the emergence of a long-range interaction in the EH, as shown in} Refs.~\cite{wang2024sudden,li2023relevant}. At the same time, the coupling decays much more slowly with distance.

\vspace{\baselineskip}
\noindent{\bf Discussion}\\
\noindent\textit{\color{blue}The advantage and disadvantage.-}
We have to emphasize that the computational complexity of this method is exponential/polynomial for {the number of degrees of freedom in $A$/$B$}. Since we trace all the degrees of freedom of $B$ and sample them, the complexity of $B$ part is as same as the normal QMC which increases in the power-law. At the same time, we sample all the degrees of freedom of $A$ without any trace operation, thus its complexity is proportional to the dimension of the RDM, which is exponentially increasing. Therefore, the advantage of this method is that it is basically not limited to the size of the environment and 
{it enables the extraction of both} the RDM and fine levels of the ES. 

{However, {since} the important information is always contained in the low-lying ES, {and} the importance sampling of QMC just presents the low-lying levels first, actually we can set a cut-off for the sampling to avoid the additional cost for high levels. 
{For example, focusing on the first excited gap of the ES (the Schmidt gap\cite{bayat2014order,PhysRevB.97.201105}) could significantly reduce the required sampling.}
As shown in the Fig.~\ref{Fig_MC_ED}(a), the first gap converges very quickly {during} the Monte Carlo sampling.}

To compare with ED, this method can obtain $\rho_A$ 
{for much larger system sizes}. To compare with the QMC+SAC scheme, it is not {restricted} to the cut-geometry between $A$ and $B$, and it can extract the fine levels of low-lying ES. (In the QMC+SAC method, because the spectral function is gained from the imaginary time correlation function via numerical analytic continuation, it requires that $A$ should have periodic boundary condition to guarantee a well-defined momentum $k$.) 
In other words, the analytic continuation depends closely on the cut form of the entanglement. However, our method can study any geometry for the $A$. Similarly, although the Bisognano-Wichmann theorem gives the detailed form of entanglement Hamiltonian, it holds under some strict conditions (e.g., the entanglement region has no corner, the original Hamiltonian satisfies translation symmetry) and may lose the effectiveness in many lattice models~\cite{dalmonte2022entanglement,mendes2019entanglement,Mendes-Santos_2020,giudici2018entanglement}.

\noindent\textit{\color{blue}Conclusions and outlooks.-}
We propose a practical scheme to extract the high-precision RDM (displayed in fine levels of entanglement spectrum) from the method of quantum Monte Carlo simulation combined with exact diagonalization. It can extract the fine information of low-lying ES for large-scale quantum many-body systems whose environment has huge degrees of freedom. Using this method, we explain the previous discrepancy for the entanglement spectrum in a Heisenberg ladder system. Furthermore, we calculate the ES of a 2D AFM Heisenberg model within two different cut-geometries. Both the cornered and cornerless cases show the TOS character which reflects the continuous symmetry breaking here. {Moreover, we have put forward an efficient and simple way to restore the entanglement Hamiltonian from the data of sampled RDM.}

In summary, the reduced density matrix can be obtained in larger size via QMC within this simple frame. Therefore, other observables such as von Neumann entropy, entanglement negativity and off-diagonal measurements can also be extracted in this way~\cite{wang2024entanglementmicroscopy}. This low-technical-barrier algorithm can be widely used in quantum Monte Carlo simulations and is easy to be parallelized on computer.

\vspace{\baselineskip}
\noindent{\bf Methods}

We use quantum Monte Carlo algorithm combined with exact diagonalization technique in this work. Details have been explained in the main text.

\vspace{\baselineskip}
\noindent{\bf Data availability}

The data that support the findings of this study are available at \href{https://doi.org/10.5281/zenodo.14879371}{https://doi.org/10.5281/zenodo.14879371}

\vspace{\baselineskip}
\noindent{\bf Code availability}

All numerical codes in this paper are available from the authors.

\bibliography{ES}

\vspace{\baselineskip}
\noindent{\bf Acknowledgements}
We would like to thank Zhentao Wang, Chen Cheng, Wei Zhu, Xue-Feng Zhang and Han-Qing Wu for fruitful discussions. SH acknowledges funding from MOST Grant No. 2022YFA1402700, NSFC No. U2230402, NSFC Grant No. 12174020.
BBM acknowledge the Natural Science Foundation of Shandong Province, China (Grant No. ZR2024QA194) and NSFC Grant No. 12247101. We thank the Scientific Research Project (No.WU2024B027) and the Start-up Funding of Westlake University. The authors also acknowledge the HPC centre of Westlake University and Beijng PARATERA Tech Co.,Ltd. for providing HPC resources.

\vspace{\baselineskip}
\noindent{\bf Author Contributions} 
ZY initiated the work and designed the QMC+ED algorithm. SH proposed the scheme to restore entanglement Hamiltonian. BBM performed all the computational simulations. YMD and ZW contributed to the analysis of the results, coding and writing the manuscript. SH and ZY supervised the project.

\vspace{\baselineskip}
\noindent{\bf Competing interests}
The authors declare no competing interests.\\

\end{document}